\documentclass[twocolumn]{aastex63}
\usepackage{graphicx}
\usepackage{natbib} 
\usepackage{array}
\usepackage{times}
\usepackage{amsmath}
\usepackage{xcolor}

\def\araa{{ARA\&A}}          
\def\apj{{ApJ}}                 
\def\apjl{{ApJ}}                
\def\apjs{ {ApJS}}               
\def\ao{ {Appl.~Opt.}}           
             
\def\aap{ {A\&A}}

\def\mnras{ {MNRAS}}

\def\prd{ {Phys.~Rev.~D}}

\newcommand{\msun}{{$M_{\odot}$}}
\newcommand{\mstar}{{$M_{\star}$}}
\newcommand{\mbh}{{$M_{\rm BH}$}}

\newcommand{\gtsima}{$\; \buildrel > \over \sim \;$}
\newcommand{\ltsima}{$\; \buildrel < \over \sim \;$}
\newcommand{\prosima}{$\; \buildrel \propto \over \sim \;$}
\newcommand{\gsim}{\lower.5ex\hbox{\gtsima}}
\newcommand{\lsim}{\lower.5ex\hbox{\ltsima}}
\newcommand{\simgt}{\lower.5ex\hbox{\gtsima}}
\newcommand{\simlt}{\lower.5ex\hbox{\ltsima}}
\newcommand{\simpr}{\lower.5ex\hbox{\prosima}}

\newcommand{\cxo}{\textit{Chandra}}

\newcommand{\lx}{$L_{\rm X}$}

\begin{document}


\title{Exploring the local black hole mass function below $10^6$ solar masses}
\author{Elena Gallo}
\affiliation{Department of Astronomy, University of Michigan, Ann Arbor, MI 48109, USA}

\author{Alberto Sesana}
\affiliation{Dipartimento di Fisica ``G. Occhialini", Universit\'a\ degli Studi Milano Bicocca, Piazza della Scienza 3, I-20126 Milano, Italy}

\correspondingauthor{Elena Gallo}
\email{egallo@umich.edu}





\begin{abstract}
  The local black hole mass function (BHMF) is of great interest to a variety of astrophysical problems, ranging from black hole binary merger rates to an indirect census of the dominant seeding mechanism of supermassive black holes. In this Letter, we combine the latest galaxy stellar mass function from the Galaxy And Mass Assembly survey with X-ray-based constraints to the local black hole occupation fraction to probe the BHMF below $10^6$ \msun. Notwithstanding the large uncertainties inherent to the choice of a reliable observational proxy for black hole mass, the resulting range of BHMFs yields a combined normalization uncertainty of $\lesssim$1 dex over the $[10^5-10^6]$ \msun\ range, where upcoming, space-based gravitational wave detectors are designed to be most sensitive. 
\end{abstract}
%



\section{Introduction}

The mass distribution of black holes residing in $z$$=$0 galactic nuclei is one of the primary empirical tools to map the growth of supermassive black holes across cosmic time. At the high-mass end, above $\simgt 10^7$ \msun, it establishes the average black hole mass and anchors the evolution of actively accreting black holes and their host galaxies, via the continuity equation (see \citealt{kelly12} for a review). At the low-mass end, it is expected to carry information about the dominant mechanism through which the progenitors of today's supermassive black holes were assembled at $z\simgt 15$ \citep{woods18}, set tidal disruption event rates \citep{stonemetzger}, and discriminate between competing models for quenching star formation in dwarf galaxies \citep{silk17}. Perhaps most importantly, the Laser Interferometer Space Antenna \citep[LISA,][]{2017arXiv170200786A}, will be sensitive to the black hole mass range \mbh\ $\in [10^4,10^7]$ \msun\ \citep{klein16}, where our knowledge of the mass function -- let alone its redshift evolution -- is tentative at best.

The most common approach to building the BHMF (see \citealt{vika12,davis14,shankar16,mutlu16} for recent references) relies on the assumption that massive black holes are ubiquitous in galactic nuclei, and uses an empirically-established observational proxy for the black hole mass, such as the host galaxy bulge stellar velocity dispersion, mass, or luminosity, along with an assumed (or measured) distribution for the hosts observable(s) (see \citealt{kh13} for a comprehensive review on scaling relations). Several uncertainties limit our ability to extend this kind of investigation reliably below \mbh$\simlt 10^6$ \msun\ (see \citealt{greeneho07} for a pioneering work based on low-luminosity AGN). To start with, as illustrated by, e.g., \cite{shankar13} (see also \citealt{tundo07} and \citealt{lauer07}), different choices of \mbh-scaling relations yield sizable differences in the resulting BHMF. The problem is further exacerbated at the low mass end, where a variety of seemingly unrelated issues, including (but not limited to) morphology or activity-dependent substructure (whereby, e.g., AGN vs. inactive; bulged vs. bulgeless vs. pseudo-bulged galaxies occupy different regions of the parameter space for a given \mbh\ relation; see \citealt{graham16} and \citealt{vdb16} for contrasting views) and observational biases \citep{bernardi07,shankar16} all conspire to increase the inferred scatter and make it exceedingly hard to identify a single scaling relation that can be reliably applied over a broad range of masses, luminosities or stellar velocity dispersions. To add to the controversy, most of the published works do not report on black hole mass upper limits, partly due to the known bias against publishing ``negative" results; this kind of bias is hard if not impossible to account for, and yet is all but guaranteed to affect the resulting best-fitting parameters.

A second, equally problematic issue is the so-called black hole occupation fraction, defined as the fraction of galaxies that host a nuclear massive black hole, irrespective of our ability to detect it, or weigh it. Whereas there is general agreement that massive black holes are ubiquitous at higher host galaxy stellar masses (\mstar$\simgt 10^{11}$ \msun), the question remains open down the mass function, where the smaller gravitational sphere of influence and/or lower accretion luminosity make it more arduous to infer (or rule out) the presence of a black hole. An additional complication is the possibility that black holes in dwarf galaxies may not be bound to reside in their nuclei, posing further challenges to any systematic observational search \citep{bellovary19}.  
Lastly, obtaining a deep and complete luminosity/mass function for the host galaxies has not been possible until fairly recently.
These limitations have henceforth hampered any quantitative effort to explore the BHMF below $\simlt 10^6$ \msun. Here, we make a first attempt in this direction by (i) leveraging the full extent of the galaxy stellar mass function delivered by the Galaxy And Mass Assembly (GAMA) survey data \citep{gama}, and (ii) including observationally-driven constraints on the local black hole occupation fraction in the host stellar masses regime below $\simlt 10^{10}$ \msun. 
We conclude by illustrating how sizable improvements in the local occupation fraction constraints may be impacting the determination of the BHMF over the next decades.

\section{\label{sec:of}Occupation fraction constraints}

We perform a re-analysis of the \cxo\ \textit{X-ray Observatory} imaging data for the sample of 194 nearby ($\simlt$30 Mpc) early type galaxies considered by \cite{miller15} (see their table 1), who developed a Bayesian formalism to convert the measured X-ray active fraction for a large unbiased sample with uniform luminosity coverage into occupation fraction as a function of the host stellar mass. Added to the original sample are an additional 132 early types with uniform \cxo\ coverage down to the same sensitivity, for a total of 326 targets \citep{gallo19}.  We summarize below the methodology and its key assumptions.  

Granted a robust statistical assessment of the X-ray binary (XRB) contamination to the nuclear signal, high-resolution X-ray imaging offers the cleanest diagnostics of highly sub-Eddington accretion-powered emission, effectively bridging the gap between AGN and formally inactive nuclei \citep{zhang09,desroches09,gallo10,grier11,miller12,lemons15,foord17,she17}. The \cxo\ observations for this sample achieve a uniform (0.5-8 keV) luminosity threshold of $\log$\lx$=38.3$ (in CGS units), hence probing highly sub-Eddington ratios for massive BHs. The data also indicate a quantitative scaling between the nuclear X-ray luminosity, \lx, and the host galaxy stellar mass, \mstar\ (which, presumably, stems from an underlying relationship between black hole mass and host properties). Given the existence of such a correlation, and modulo a reliable XRB contamination assessment, the efficiency with which an actively accreting, massive black hole is detected is set by the intrinsic fraction of galactic nuclei that are occupied by a massive black hole. To infer the occupation fraction, we first generate a population of 10,000 galaxies whose stellar mass distribution is constructed to match the target galaxies' distribution. These are populated with nuclear black holes according to an analytical prescription where the probability of hosting a black hole is given by:
\begin{equation}\label{eq:of}
\lambda_{\rm occ}(M_{\star})=
0.5+0.5 \tanh\bigg( 2.5^{|8.9-\log{M_{\star,0}}|}\log{\frac{M_{\star}}{M_{\star,0}}  }\bigg ),
        \end{equation}
with occupation fractions bounded by $\lambda_{\rm occ}\simeq 0$ for
$M_{\star}<10^{7} M_{\odot}$ and $\lambda_{\rm occ}\simeq1$ for $M_{\rm
  star}>10^{10} M_{\odot}$. 
\begin{figure}
\begin{center}
  \includegraphics[width=\linewidth]{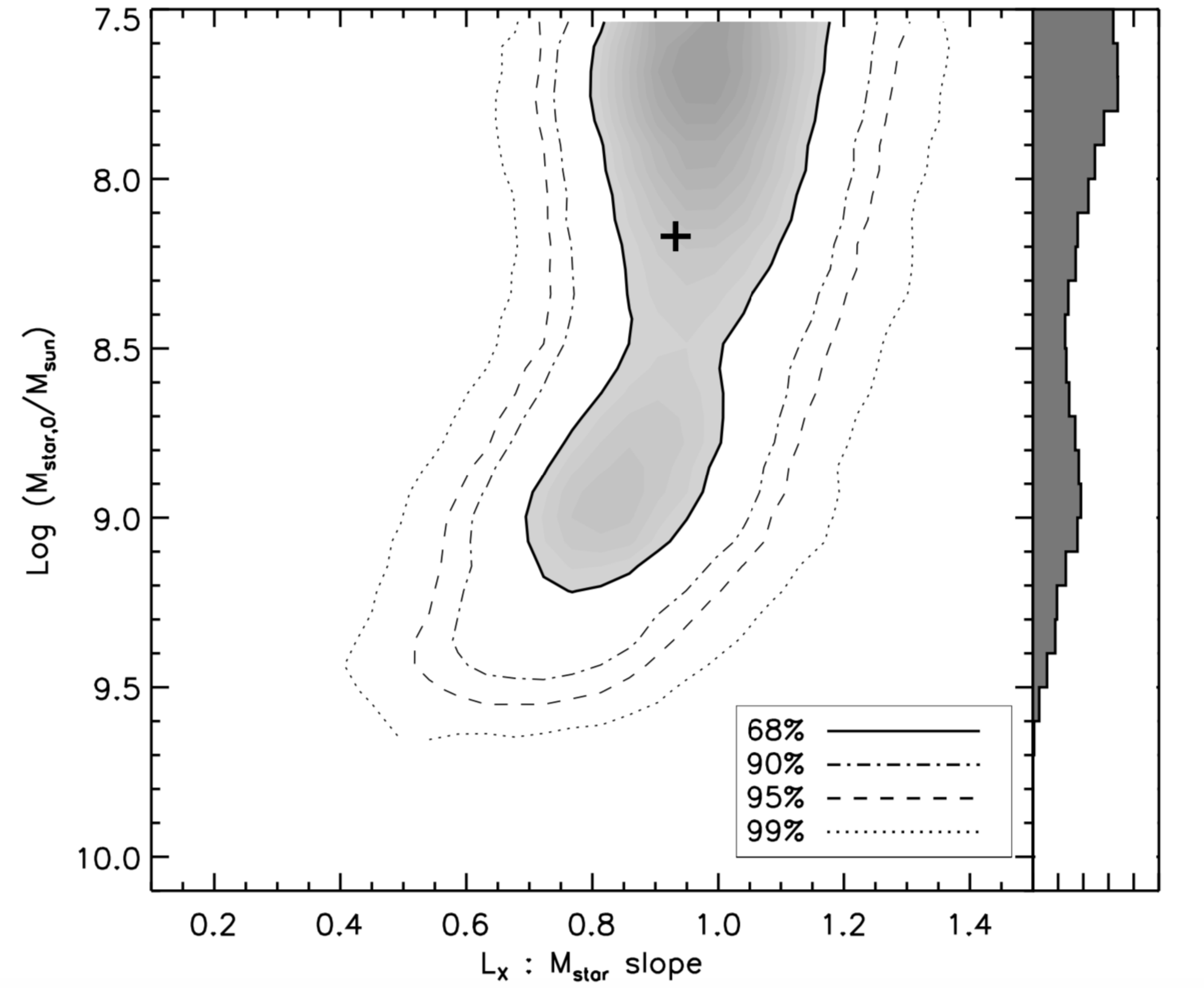}
  \end{center}
 \caption{Updated constraints on the black hole occupation fraction via the $M_{\star,0}$ parameter in Equation \ref{eq:of}, based on 326 nearby early type galaxies with uniform \cxo\ coverage. Shown here is the posterior distribution of $M_{\star,0}$ vs. the slope of the \lx:\mstar\ relation, with the median values shown by a cross. }
 \label{fig:of}
\end{figure}
Each galaxy is then assigned a nuclear luminosity on the basis of on the best-fitting \lx:\mstar\ correlation, under the (arbitrary) assumption that the degree of intrinsic scatter remains constant across the stellar mass range. The actual luminosity for each galaxy stems from (i) imposing a sensitivity threshold that matches the data's, and (ii) modifying the assigned value (after accounting for intrinsic scatter) to an upper limit for all galaxies that either lack a black hole, or have a black hole that is emitting below the set sensitivity threshold. XRB contamination is incorporated into the analysis by fitting each distribution $1,000$ times and probabilistically varying whether each X-ray source is treated as a detection or a limit, according to the estimated probability that the nuclear X-ray emission is associated with a black hole as opposed to an XRB (see \citealt{lee19}, and references therein, for a detailed description of the XRB assessment). 

The best-fitting, preferred values for the four parameters under consideration (i.e.; $M_{\star,0}$, plus the \lx:\mstar~ relation slope, intercept and scatter) are taken as the median of 5,000 (thinned from 50,000, retaining every tenth) draws from the posterior distribution. 
Figure \ref{fig:of} shows the posterior distribution of $M_{\star,0}$ vs. the slope of the \lx:\mstar\ relation. In terms of occupation fraction, this corresponds to values higher than 47\% for host stellar masses below $10^{10}$ \msun\,  (68\% C.L.); values lower than 27\% are ruled out with 99.99\% confidence.
The $M_{\star,0}$ posterior distribution shown in the right-hand outset is modeled as the sum of two (truncated) Gaussians and folded in the BHMF expression through the occupation probability $\lambda_{\rm occ}$, as described below. 

\section{\label{sec:mf} Mass function modeling}

The BHMF, $\Phi$, is defined as the number of black holes per comoving volume $V(z)$ in the mass interval [\mbh, \mbh$+d$\mbh]:
\begin{equation}
\Phi(M_{\rm BH},z)=N\left(\frac{dV}{dz}\right)^{-1}p(M_{\rm BH},z)
\end{equation}
where $p$ is the joint probability distribution of \mbh\ and $z$. The integral of the BHMF over volume and redshift yields its normalization $N$, i.e., the total number of black holes in the observable universe. Generally speaking, estimating the BHMF based on the observed distribution of black hole mass estimates requires a careful assessment of the distribution of the observable quantities (typically, flux) which define the selection function of the chosen sample, to correct for incompleteness \citep{kelly09,kelly12,shankar13}. In the case where the masses are derived directly from a second observational quantity $y$ whose distribution $\Phi$ is known, the expression for the BHMF function simplifies to $\int p(M_{\rm BH},z)\Phi(y)dy$. This approach is usually followed to derive the BHMF at $z$ $=$0, where the distribution of black hole masses follows from (one of) the local scaling relation(s) with a specific host galaxy property. 
With this in mind, we express the local BHMF as: 
\begin{equation}\label{eq:mf}
\begin{split}
\Phi (M_{\rm BH}) =\int \Phi(M_\star)\lambda_{\rm occ}(M_\star) \frac{1}{\sqrt{2\pi \eta^2}}  \\
 \exp {\left[- \frac{M_{\rm BH} - (\alpha+\beta M_\star)^2}{2\sigma^2} \right]dM_\star
},
\end{split}
\end{equation}
where $\Phi(M_\star)$ is the galaxy stellar mass distribution; $\alpha$, $\beta$ and $\sigma$ are the intercept, slope and scatter of the functional relationship between \mbh\ and the \textit{total} host galaxy stellar mass, \mstar; and $\lambda$ is the occupation fraction probability defined above.

The functional parametrization for the galaxy stellar mass function is taken from \cite{gama}, which makes use of the full GAMA data set II to extend the stellar mass function down to  $10^{7.5}$ \msun. For the purpose of this work, we simply note that, below the well known upturn at $\sim$$10^{9.5}$ \msun, the distribution appears to follow a single power-law. 
\begin{figure}
 \includegraphics[width=\linewidth]{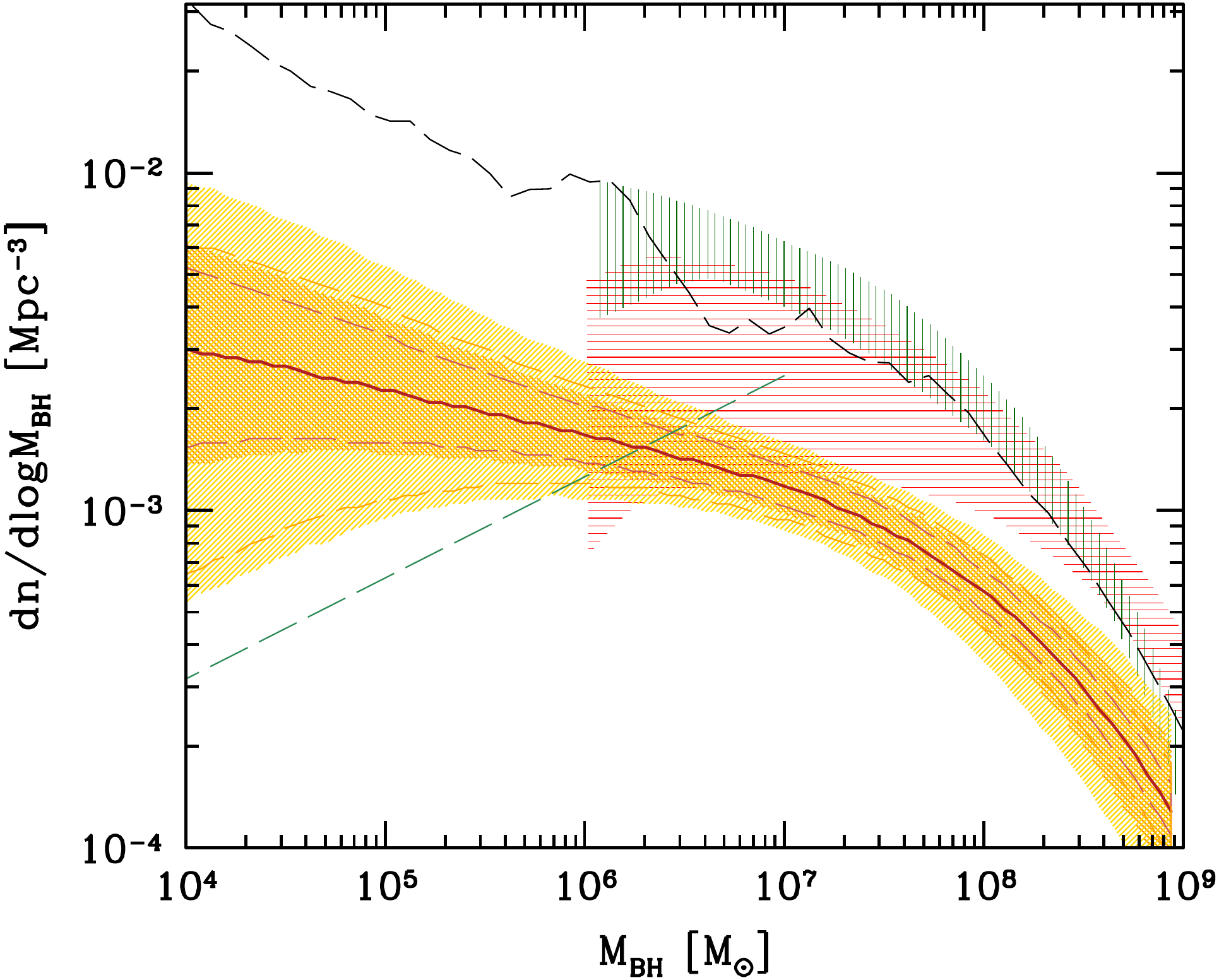}
 \caption{The occupation-corrected BHMF defined by Equation \ref{eq:mf}. The solid red line traces the median; the dark and light yellow areas trace the 68 and 95\% confidence regions, respectively, with all systematic errors included (whereas the long-dashed dark and light yellow lines trace the same confidence level regions ignoring systematic errors). This BHFM is derived from Equation \ref{eq:mf}, where the adopted observational proxy for black hole mass is the host galaxy total stellar mass, and the best-fitting parameters for the scaling relation are derived for the AGN+inactive galaxy sample assembled by \cite{rv15}. For comparison, the hatched green and red areas above $10^6$ \msun\ are taken from \cite{shankar09} and \cite{shankar16}, respectively; the long-dashed gray line is from \cite{barausse12} and the long-dashed green line is from Gair et at. (2010; see main text for details).}\label{fig:mf}
\end{figure}
For the scaling relation with black hole mass, we adopt a similar stance to \cite{rv15}, who investigate a relationship with total stellar mass for two heterogeneous, nearby samples; $\sim$$80$ local, inactive (elliptical, S and S0) galaxies with dynamical black hole mass estimates, plus $\sim$$260$ nearby ($z$$<$$0.055$) type I AGN (including a small sub-sample of dwarf galaxy hosts), for which black hole masses are obtained through either single-epoch virial scaling relations, reverberation mapping, or dynamical modeling.

Although \cite{rv15} conclude that a single linear relation between $\log$(\mbh) and $\log$(\mstar) is disfavored by the data, and proceed to fit the AGN and inactive galaxy samples separately, we take a more crude approach and consider the full AGN plus inactive galaxy sample (for reference, running either a Kendall-$\tau$ or Spearman test suggests a moderately strong correlation for the combined AGN plus inactive galaxy sample, with rank correlation parameters $\tau=0.38$ and $\rho=0.53$, respectively). 
This choice is motivated by several issues which plague the reliability of the known scaling relations in the low mass regime. Adopting a large-scatter relation such as the \mbh:\mstar\ -- albeit intrinsically questionable -- serves the purpose of \textit{bracketing the large uncertainties spanned by other relations}, and in particular that with stellar velocity dispersion, whose functional shape becomes highly controversial for low mass objects \citep{kh13,vdb16,kormendy16,shankar16,shankar19}.

\begin{figure*}[htp]
\begin{center}
 \includegraphics[width=0.485\linewidth]{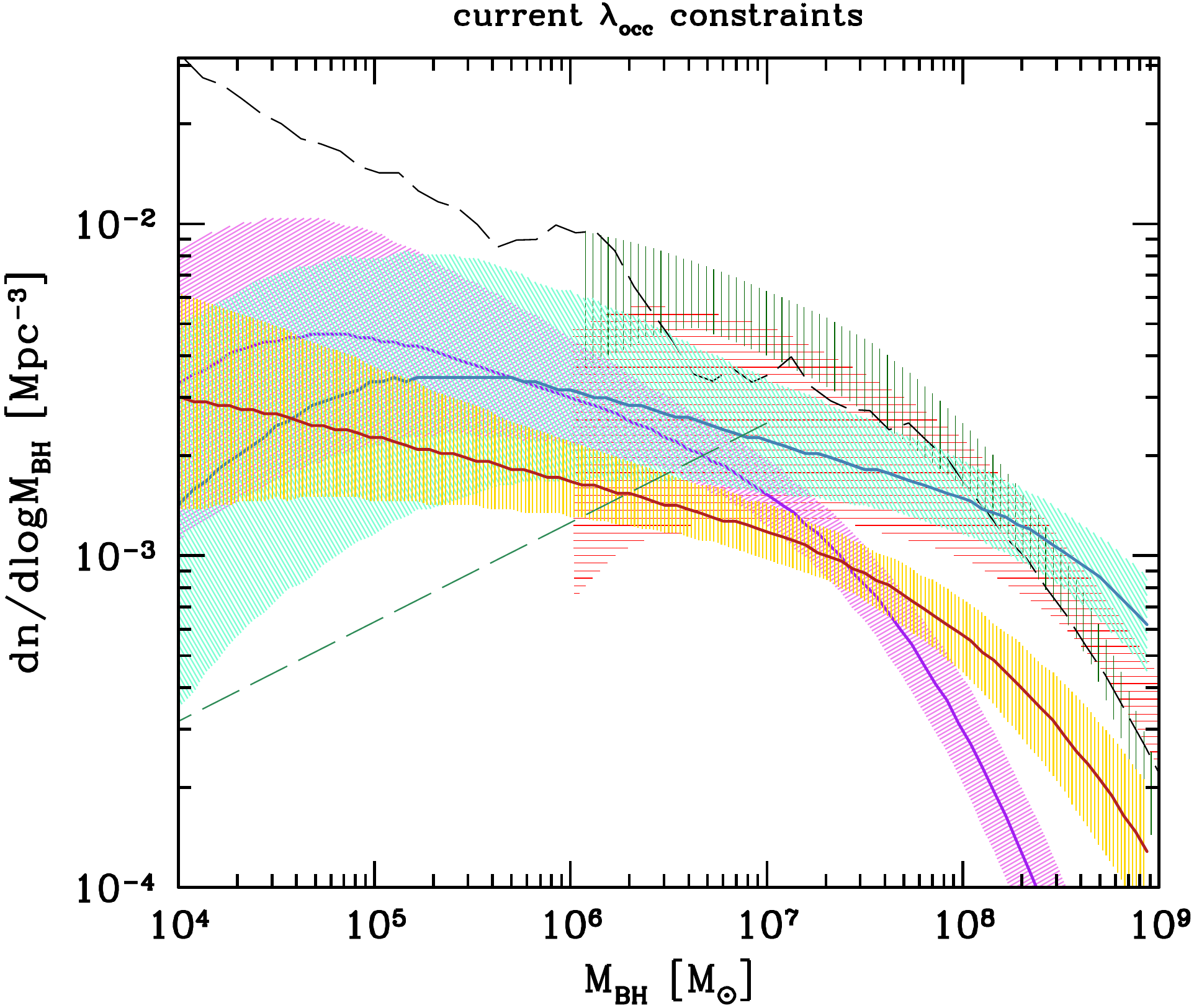}
  \includegraphics[width=0.485\linewidth]{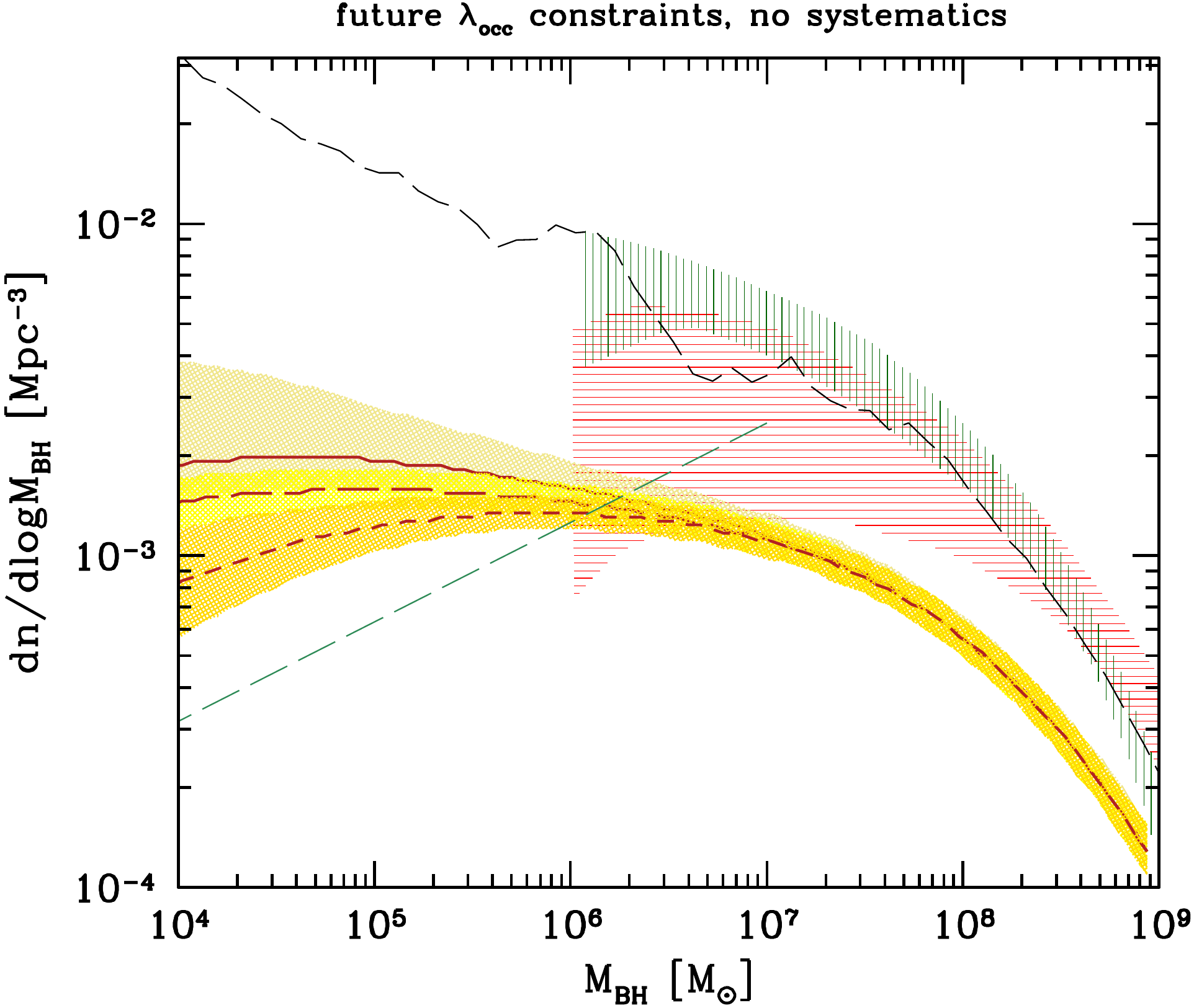}
 \caption{Left: Illustration of how different choices of black hole mass scaling relations affect the inferred BHFM. The yellow curve covers the same 68\% confidence region as shown in Figure \ref{fig:mf}, i.e., based on the \mbh:\mstar\ scaling relation that we obtained by fitting the whole AGN + inactive galaxies sample assembled by \cite{rv15}. For comparison, the purple and blue curves are derived using the scaling relation derived by \cite{rv15} by fitting the AGN and inactive galaxy samples, respectively. All other lines and areas have the same meaning as in Figure \ref{fig:mf}. Right: How achieving a few per cent accuracy in the black hole occupation fraction determination could shrink the errors in the ensuing BHMF. These curves are derived adopting simulated posterior distributions for $M_{\star,0}$, that would result from assembling X-ray imaging data for 5,000 (as opposed to $\simeq 300$) nearby ($\simlt $100 Mpc, as opposed to $\simlt 30$ Mpc) galaxies with uniform X-ray coverage and sub-arcsec spatial resolution (adapted from \citealt{gallo19}). Shown in yellow are the results from three realizations with ``true" $\log (M_{\star,0}/M_{\odot})$ values of 8.25, 8.75 and 9.25 (from top to bottom), and adopting the same \mbh:\mstar\ relation that yields the yellow curve in the left panel.}
 \label{fig:mfs}
 \end{center}
\end{figure*}
We explore a relation of the form  $\log$(\mbh/\msun) $=\alpha+\beta \log$(\mstar/$10^{11}$\msun), with intrinsic scatter, $\sigma$, included in the fit. To do so, we use the linear regression analysis developed by \cite{kelly07} and implemented in \textsc{IDL} as \textsc{linmix\_err}. For easiness of comparison, we adhere to the same choice of errors as \cite{rv15}, both in terms of measurement errors as well as quoted uncertainties in the parameters' best-fitting posterior distribution. For the full AGN plus inactive galaxy sample, we obtain $\alpha=8.13\pm0.09$, $\beta=1.72\pm0.14$ and $\sigma=0.61\pm 0.05$. Not surprisingly, the scatter is significantly larger than what obtained by \cite{rv15} for the individual samples (i.e., 0.24 and 0.47 dex, for the AGN, and inactive galaxies, respectively).

The resulting, occupation-corrected BHMF is shown in Figure \ref{fig:mf}, in addition to: (i) the BHMFs derived by \cite{shankar09} and \cite{shankar16} (respectively hatched green and hatched red areas above $10^6$ \msun); (ii) results from semi-analytical models by \cite{barausse12}, combining heavy and light black hole seeds (long-dashed gray line); (iii) the functional prescription by \cite{gair10}, who assume a BHMF that declines with decreasing mass as \mbh$^{0.3}$ below $10^6$ \msun\ (long-dashed green line). For comparison, the left panel of Figure \ref{fig:mfs} illustrates the BHMFs obtained by adopting the AGN (purple) and inactive galaxies (blue) \mbh:\mstar\ relations obtained by \cite{rv15}. As expected, fitting the entire sample yields a mass function (shown in yellow) that broadly encompasses either, with obvious differences both at low and high masses.

For replicability purposes, the median (solid red line) of our best-guess BHMF in Figure \ref{fig:mf} can be well-approximated by the Schechter-like function:
\begin{equation}
\log \Phi(M_{BH})=c_1+c_2\log(M_{\rm BH}/M{_{\odot}})+c_3(M_{\rm BH}/M{_{\odot}}),
\end{equation}
with coefficients $c_1=-2.13$, $c_2=-0.098$ and $c_3=-0.00011$ (see \citealt{schechter}).

\section{Discussion}\label{sec:disc}

Overall, the main factor driving the uncertainty in the expected BHMF remains the choice of the observational proxy for \mbh. This highly contentious issue will hopefully see a resolution in the upcoming era of $\simgt$30 meter-class ground-based telescopes, which promise to usher a complete and unbiased census of the black hole population in the local volume down to the dwarf galaxy regime  \citep{greene19}. 
In the meantime, as a proxy for \mbh\ we have chosen to adopt a  large-scatter scaling relation with \mstar, so as to embrace the range of uncertainties spanned by other (likely more fundamental) relations at low masses, and with the implicit understanding that, \textit{at high masses, the resulting BHMFs are markedly less reliable than existing estimates} (which typically adopt a scaling relation with stellar velocity dispersion as a proxy for \mbh). 

As expected, the additional implementation of an occupation fraction correction has a non-negligible impact at low masses. Its effect, however, is more readily apparent for shallower (i.e., having lower slope values, $\beta$) and/or higher normalization (higher $\alpha$) \mbh:\mstar\ relations, such as those established by the AGN and inactive galaxy samples in \citealt{rv15} (having, respectively, $\beta=1.05$, $\alpha=7.45$, and  $\beta=1.4$, $\alpha=8.95$). The BHMFs arising from those relations are shown in purple (AGN) and blue (inactive galaxies) in Figure \ref{fig:mfs}, and compared to the combined-sample relation BHMF, again in yellow. The combination of slope and intercept for both sub-sample relations works in such a way that, at low masses, \mbh\ for a given \mstar\ is lower for the combined-sample relation than it is from either the AGN or inactive galaxy sample relation. For the same $M_{\star,0}$ value, thus, galaxies with low occupation end up bracketing a lower \mstar\ \textit{range} than they would for either the AGN or inactive sample relation. As a result, a downturn is visible for both the purple and blue curves in Figure \ref{fig:mfs}, while it is not as obvious for the yellow curve. For similar reasons, that Shankar's BHMFs exceed our best guess BHMF (yellow curve) at high masses can be understood in terms of the underlying \mbh:velocity dispersion relation likely yielding a higher \mbh\ for a given velocity dispersion-based \mstar\ (a la Faber-Jackson) than our best guess relation.   

Overall, while the median BHMFs yielded by the three different sample relations span over 1 dex above \mbh$\simgt 10^8$, they differ only by a factor $\simlt 3$ below $\simlt 10^7$\msun. Most notably, the resulting BHMF normalization (68\% C.L.) uncertainty in the mass range $[10^5-10^6]$ \msun, where the LISA sensitivity peaks, is less than one order of magnitude, irrespective of the scaling relation choice. These values are intermediate between the optimistic (LISA-wise) estimates by Barausse et al (2012), and the most pessimistic prescription, by Gair et al. (2010), which represent respectively the highest and lowest normalization BHMFs adopted by \cite{babak17} towards their predictions for extreme mass ratio inspiral event rates for LISA. \\

We further emphasize that the occupation fraction constraints employed throughout this work are based on a sample of early type galaxies. While this choice is chiefly driven by feasibility reasons, and more specifically the goal of minimizing contamination from bright high mass XRBs to the nuclear signal, it represents a serious drawback for our estimates, especially considering the increased prominence of late types at low stellar masses \citep{blanton09}.  
Although nuclear XRB contamination in star forming galaxies is arguably tractable problem, the lack of a comparably large, unbiased sample with uniform \cxo\ coverage is the main limiting factor that prevents us from extending a similar investigation to late types at the present time. 

To partially account for this shortcoming, we explore how highly improved occupation fraction constraints at low \mstar\ might affect the resulting BHMF. This is done by leveraging a suite of simulations aimed at establishing the number of galaxy targets that are necessary to achieve a few per cent accuracy in occupation below \mstar$\simlt 10^{10}$ \msun\ \citep{gallo19}.
In general, conservative occupation fraction values can be recovered with $\sim$5$\%$ accuracy with $\sim$5,000 galaxies (adopting an \mstar-dependent sensitivity reduces the number of required targets to $\sim$3,000; Hodges-Kluck et al., submitted). This can be realistically achieved by an X-ray instrument with large collecting area and sub-arcsec imaging capabilities over a wide-field of view, such as those envisaged for the \textit{Lynx} X-ray Surveyor \citep{Gaskin18} mission, or the Advanced X-ray Imaging Satellite \citep{mushotzky18} probe.  The effect of a highly improved occupation accuracy on the BHMF is illustrated by the right panel of Figure \ref{fig:mfs} for three ``pessimistic" input values of $M_{\star,0}$ (i.e., all implying lower occupation than the current median value of 8.17, shown in Figure \ref{fig:of} as a cross). If a single, reliable scaling relation with constant scatter were to be established down to such low host galaxy masses, the corresponding BHMF uncertainty would shrink to $\lesssim 2$. \\

In conclusion, notwithstanding the large uncertainties inherent to the choice of a reliable scaling relation with \mbh, once an observationally-motivated occupation fraction correction is implemented at the low mass end, the resulting range of BHMFs yields a combined 68\% normalization uncertainty of $\lesssim$1 dex in the $[10^5-10^6]$ \msun\ range, where upcoming gravitational wave space detectors such as LISA will be most sensitive. 

\acknowledgments 
This research was supported in part by the National Science Foundation under Grant No. NSF PHY-1748958. We wish to thank the Kavli Institute for Theoretical Physics, the Munger Physics Residence, and the Rice Family Fund for the generous hospitality and support during the development of this work.

\end{document}